\documentclass[preprint,3p,times]{elsarticle}
\usepackage{graphicx}
\usepackage[utf8]{inputenc}
\usepackage{bm}
\usepackage{amsmath}
\usepackage{amsfonts,amssymb}
\usepackage{color}
\usepackage{xcolor}
\usepackage[T1]{fontenc} 
\usepackage[normalem]{ulem}
\usepackage{float}
\usepackage{soul}
\usepackage{amssymb}

\journal{Physics Letters B}

\begin{document}
\begin{frontmatter}

\title{On the Generalized Uncertainty Principle and Cosmology.}

\author{Oscar L\'opez-Aguayo$^a$}
\ead{o.lopezaguayo@ugto.mx}
\author{J. C. L\'opez-Dom\'inguez$^{a,b}$}
\ead{jlopez@fisica.uaz.edu.mx}
 \author{M. Sabido$^a$}%
\ead{msabido@fisica.ugto.mx}
\address{ 
$^a$ Departamento  de F\'{\i}sica de la Universidad de Guanajuato,\\
 A.P. E-143, C.P. 37150, Le\'on, Guanajuato, M\'exico.\\
 $^b$Unidad Acad\'emica de F\'isica, Universidad Aut\'onoma de Zacatecas, Calzada Solidaridad esquina con Paseo a la Bufa S/N C.P. 98060, Zacatecas, M\'exico.\\
}%
\begin{abstract}
In this work we study the effects of the generalized uncertainty principle (GUP) in cosmology. We start with the Friedmann-Robertson-Walker (FRW) model endowed with a scalar field. After introducing the GUP modification to the model, we solve for the quantum and classical cases. Finally we find the GUP modified Friedmann equations.
\end{abstract}
\end{frontmatter}
\section{Introduction}

The $\Lambda$CDM cosmological model is the best phenomenological description of the observed Universe. It is constructed using General Relativity (GR) with the standard model of particles.   Although the $\Lambda$CDM model is compatible with observations, as it is a  phenomenological model, it is not surprising that several theoretical aspects remain.  In particular,
the dark energy problem, where proposing a cosmological constant $\Lambda$, there are inconsistencies with traditional quantum field theory \cite{weinberg,lambda} (these problems have been addressed by different approaches \cite{polchinski}). In particular, scalar fields have been successful as an alternative for the description of dark energy \cite{SF1,ratra,SF2,SF3}. 
For example, phantom fields (scalar fields with a negative kinetic term) have been considered in the literature, these fields provide an effective negative pressure and a repulsive effect, that can be responsible for the late time accelerated expansion \cite{Caldwell}. Alternatively, we can consider the possibility that the problems related to $\Lambda$ can be  consequence of the poor understanding of gravity and modifications to gravity should be considered.

The search for a theory of quantum gravity has been one of the most emblematic and tough problem in fundamental physics. Although, until this day we do not have a consistent theory for quantum gravity several candidates has been proposed each one with its respective success. One feature that we can expect from a quantum theory of gravity, (it arises in string theory  \cite{strings}, loop quantum gravity \cite{lqg} and noncommutative gravity \cite{ncg}), is the existence of a minimal measurable length. 
{The Uncertainty Principle is at the core of quantum theory and can be considered as inherent limit for measurement in phase space. In order to introduce an uncertainty in the positions (and therefore a corresponding minimum length) modifications to Heisenberg's uncertainty principle have been proposed, this is known as the Generalized Uncertainty Principle (GUP). This idea has been explored in a myriad of physical contexta, quantum physics \cite{quantum}, solid state physics \cite{grafeno}, quantum optics \cite{optics} and gravitational waves \cite{gw}.}
{The GUP can have an important role in the early stages of the Universe, hence, in order to understand the present dynamics of the Universe we need to implemented it and investigate its effects.}
It {is} plausible that a minimum length can be introduced to gravity by using the GUP in the Wheeler--DeWitt (WDW) equation {of the cosmological model}. These ideas where originally explored in quantum cosmology \cite{pasqualeCOSMO} and black holes \cite{pasqualeBH}, basically introducing the GUP between  the minisuperspace variables.

The main goal of this paper, is to study {the} GUP {in a FRW with}  cosmological {model}. We will focus our interest in the phantom scalar field and introduce the GUP in the WDW--equation. Moreover, we {obtain the semi-classical solutions and} study the {effects of the GUP at} semiclassical {level} and finish by deriving the GUP Friedmann equations.

The paper is arranged as follows, in section 2 we briefly review the {FRW} scalar phantom {cosmological} model. In Section 3, we review the GUP, and apply it to the model. The quantum and classical model {solutions are obtained and} analyzed. In Section 4, we {calculate and} discuss the GUP modified Friedmann equations for scalar field cosmology. Section 5, is devoted to discussion and concluding remarks.

\section{The cosmological model.}
In this section we start by presenting the model. {Consider} the Einstein-Hilbert action with cosmological constant $\Lambda$ and a minimally coupled free phantom scalar field $\varphi(t)$,
\begin{equation}\label{1}
S_{EH}=\int \sqrt{-g}\left[\frac{1}{2\kappa}\left(R-2\Lambda\right)+\frac{1}{2} \partial_{\mu}\varphi\partial^{\mu}\varphi\right]d^{4}x,
\end{equation}
for the flat FRW metric
$
ds^2 = -N^2(t)dt^2 + a^2(t)\left[ dr^2 + r^2d\Omega^2\right],
$
the action takes the form 
\begin{equation}\label{accion}
S = {V_0}\int dt \left [   -\frac{3a \dot{a}^2}{N} - a^3\left(  \frac{\dot{\varphi}^2}{2N} + N\Lambda    \right)    \right] ,
\end{equation}   
where $a(t)$ is the scale factor, $N(t)$ is the lapse function, $V_0$ comes from integrating the spatial part and we set it equal to 1 and we {use} the units\footnote{This is equivalent to $8\pi G = 1$.} $\kappa=1$. The minus sign in the kinetic term of the scalar action is the difference between the usual scalar field and the phantom scalar field. 

The canonical Hamiltonian derived from Eq.(\ref{accion}) is
\begin{equation}\label{ham}
H=-N \left[ \frac{P_a^2}{12a} +\frac{P_{\varphi}^2}{2a^3} - a^3 \Lambda \right],
\end{equation}
which, with the
change of variables
\begin{equation}\label{Trans1}
 x = \frac{a^{3/2}}{\mu}\sin{\left(\mu \varphi\right)},\quad
 y = \frac{a^{3/2}}{\mu}\cos{\left(\mu \varphi\right)},
\end{equation} 
with $\mu = \sqrt{3/8}$, the Hamiltonian Eq.(\ref{ham}) can be rewritten as a sum of two decoupled harmonic oscillators Hamiltonians
\begin{equation}
H = N\left(  \frac{1}{2}P_x^2 + \frac{\omega^2}{2}x^2   \right) + N\left(  \frac{1}{2}P_y^2 + \frac{\omega^2}{2}y^2   \right), \label{Ham1}
\end{equation}
of frecuency $\omega^2 = -\frac{3}{4}\Lambda$. If we consider the usual scalar field, the Hamiltonian is transformed into a ``ghost oscillator'' which is simply a difference of two harmonic oscillators hamiltonians \cite{huicho}. {For the purpose of this paper} the Hamiltonian in Eq. (\ref{ham}) is our starting model to deform using a GUP.

Using the canonical quantization, we obtain the corresponding WDW--equation, 
 $\hat{H} \psi = 0$, {of the model}. In the position-space representation, $P_x=-i\hbar\partial/\partial x$ and $P_y=-i\hbar\partial/\partial y$ in such a way that the WDW--equation is
\begin{align}\label{WDW}
    \hat{H}\psi(x,y)=-\frac{\hbar^2}{2}\nabla^2\psi(x,y)+\frac{\omega^2}{2}\left(x^2+y^2\right)\psi(x,y)=0,
\end{align}
where $\nabla^2$ is the Laplacian operator in 2--dimensions.

We can write the wave function with {the} eigenstates common to both operators, {with} every eigenstate uniquely specified by the quantum numbers $(n,m)$ . 

{Applying a} WKB type approximation we {can obtain} the same classical {solutions} from the 
classical limit applied to the WDW--equation.  Following the standard procedure {we propose} a wave function of the form 
\begin{equation}
\psi(x,y)\propto e^{\frac{i}{\hbar}\left(S_1(x)+S_2(y)\right)},
\end{equation}
which upon substitution in the WDW--equation Eq.\eqref{WDW} in the limit $\hbar\rightarrow 0$, with the approximations
\begin{equation}\label{wkb}
\left( \frac{\partial S_1(x)}{\partial x} \right)^2 \gg \frac{\partial^2 S_1(x)}{\partial x^2},~~~ \left( \frac{\partial S_2(y)}{\partial y} \right)^2 \gg\frac{\partial^2 S_2(y)}{\partial y^2},
\end{equation}
we get the Einstein-Hamilton-Jacobi equation (EHJ)
\begin{align}
& \left( \frac{\partial S_1(x)}{\partial x} \right)^2 + \left( \frac{\partial S_2(y)}{\partial y} \right)^2 + \omega^2(x^2 + y^2)= 0.
\end{align}
From the standard identification $\dfrac{\partial S_1(x)}{\partial x} = P_x$, $\dfrac{\partial S_2(y)}{\partial y} = P_y$ {, and the relation between}
$P_x,P_y$ {and $\dot{x}$, $\dot{y}$} from $\dot{x} = \left\{ x,H \right\}, \dot{y} = \{ y,H \}$, the EHJ equation becomes
\begin{equation}
\dot{x}^2 + \dot{y}^2+ \omega^2\left( x^2 +y^2\right) = 0,
\end{equation}
this equation  is equivalent to the equations of motion that are derived from the Hamiltonian, and consequently have the same solutions. 
 
\section{The GUP Cosmological Model}
As the GUP is {initially proposed} in quantum mechanics, we will introduce the GUP on the WDW--equation of our cosmological model. There are different 
approaches for the GUP, we will consider  two implementations. The first one implies a deformation on the momentum, while the second one is a deformation on the coordinates.

We start with the following by introducing the GUP
\begin{equation}\label{35}
\left[q_i,P_j\right]=i \hbar\delta_{ij} \left(1-2\beta\gamma P_0+4\epsilon\gamma^{2}P_0^{2}\right),
\end{equation}
where $P_0^2=P_{0j}P_{0j}$, { $\beta$, $\epsilon$ and $\gamma$ are constants with convenient units,} and the canonical variables {$q_i$ and $p_i$} satisfy
\begin{equation}\label{q}
q_i=q_{0i},\hspace{.7cm}
P_i=P_{0i}\left(1-\beta\gamma P_{0}+2\gamma^{2}\frac{\beta^{2}+2\epsilon}{3}P_{0}^{2}\right),
\end{equation}
{and the variables $q_{0i}$ and $P_{0j}$ satisfy the usual relation $\left[q_{0i},P_{0j}\right]=i\hbar\delta_{ij}$.}
{Substituting the relations Eq.\eqref{q} in Eq.\eqref{Ham1}}, we construct the GUP modified Hamiltonian, 
\begin{eqnarray}\label{HGUP}
    H_{GUP}&=&H_{0}-\beta\gamma\left(P_{0x}^{2}+P_{0y}^{2}\right)^{3/2}+\gamma^{2}\left(P_{0x}^{2}+P_{0y}^{2}\right)^{2}\left(\frac{\beta^{2}}{6}+\frac{2\epsilon}{3}\right)+2\beta\gamma^{3}\left(\frac{\beta^{2}+2\epsilon}{3}\right)\left(P_{0x}^{2}+P_{0y}^{2}\right)^{5/2}\nonumber\\
    &+&\gamma^{4}\left(\frac{\beta^{2}+2\epsilon}{3}\right)\left(P_{0x}^{2}+P_{0y}^{2}\right)^{3},
\end{eqnarray}
{where $H_0$ is the unperturbed Hamiltonian in Eq.~\eqref{Ham1}}.
{We can proceed to the quantization using the ladder operators, as is  usual for the quantum oscillator.} Let us start {by considering} a perturbation to the Hamiltonian to order $\gamma^2$, 
clearly the usual {creation, annihilation and number} operators, $a$, $a^{\dagger}$ and $N$, only work with the unperturbed Hamiltonian $H_{0}$. Therefore we need a set of operators for the $\gamma$ perturbations \cite{pasquale4}. To construct the new operators
$\tilde{a}$, $\tilde{a}^{\dagger}$, $\tilde{N}$, we impose the {same} conditions {on them as} for the unperturbed case
\begin{equation}
\tilde{a} | \phi_{n}\rangle =\sqrt{n} | \phi_{n-1}\rangle, \quad\tilde{N} | \phi_{n}\rangle =n| \phi_{n}\rangle, \quad
\tilde{a}^{\dagger} | \phi_{n}\rangle=\sqrt{n+1}| \phi_{n+1}\rangle, \quad
\tilde{N}=\tilde{a}^{\dagger}\tilde{a},\quad
\left[\tilde{a},\tilde{a}^{\dagger}\right]=1,
\end{equation}
where $|\phi_{n}\rangle =| \psi_{n}^{(0)}\rangle +\gamma|\psi_{n}^{(1)}\rangle+\gamma^{2}| \psi_{n}^{(2)}\rangle+\cdots$. {$|\phi_n\rangle$ are the eigenfunctions of the perturbed Hamiltonian Eq.\eqref{HGUP}, and $\psi_n^{(0)}$ are the eigenfunctions of the unperturbed Hamiltonian, and $\psi_n^{(m)}$ are the $m$--th order correction to the eigenstate.}
Then, {if we assume that the perturbation to the Hamiltonian is} small, the general form of the operators are given by
\begin{equation}\label{GUPope}
\tilde{a}=a+\sum_{n=1}^{\infty}\gamma^{n}\alpha_{n},\quad
\tilde{a}^{\dagger}=a^{\dagger}+\sum_{n=1}^{\infty}\gamma^{n}\alpha_{n}^{\dagger},\quad
\tilde{N}=N+\sum_{n=1}^{\infty}\gamma^{n}\nu_{n},
\end{equation}
{where $\alpha^\dagger_n$, $\alpha_n$ and $\nu_n$ are the $n$--th order correction for the creation, annihilation and number operator} 
We can explicitly apply these operators {Eq.\eqref{GUPope}}to the wave function
\begin{equation}
 \tilde{a}| \phi_{n}\rangle =\left(a+\sum_{m=1}^{\infty}\gamma^{m}\alpha_{m}\right)\left(| \psi_{m}^{(0)}\rangle +\sum_{m=1}^{\infty}  \gamma^{m}| \psi_{n}^{(m)}\rangle \right),\quad
  \tilde{a}^{\dagger}| \phi_{n}\rangle =\left(a^{\dagger}+\sum_{m=1}^{\infty}\gamma^{m}\alpha_{m}^{\dagger}\right)\left(| \psi_{m}^{(0)}\rangle +\sum_{m=1}^{\infty}  \gamma^{m}| \psi_{n}^{(m)}\rangle \right),\label{a}
 \end{equation}
 {and using the procedure in \cite{pasquale4} to obtain} {we can calculate} the expansion of the operators to the desired order. {Taking $\omega=\sqrt{\frac{3\left|\Lambda\right|}{8}}$, in $H_{GUP}$, and after expanding to order $\gamma$, we get}
for the operators $\tilde{a}$,  $\tilde{a}^{\dagger}$ 
\begin{equation}
\tilde{a}=a-\frac{\beta\gamma}{2}\left(\frac{3\left|\Lambda\right|}{2}\right)^{1/4}\left(a^{\dagger 3}-6 N a^{\dagger }+2 a^3\right),\quad\tilde{a}^{\dagger}=a^{\dagger }+\frac{\beta\gamma}{4}\left(\frac{3\left|\Lambda\right|}{2}\right)^{3/4}
   \left(a^{\dagger 3}-6 N a^{\dagger}+2a^3\right),
\end{equation}
{now, we substitute these in Eq. \eqref{q}} that after substituting in the momentum and coordinates {Eq.\eqref{q}}, we get
\begin{eqnarray}
p_{0k}=i\left(\frac{3 |\Lambda|}{32}\right)^{1/4}\left.
(a^{\dagger}_{k}-a_{k}\right),~
{\tilde{q}_k=q_{0k}},~\tilde{k}=\left(\frac{2}{3\left|\Lambda\right|}\right)^{1/4}\left(a^{\dagger}+a\right),~ {p_k}={p_{0k}}+\frac{i}{8}  \left(\frac{3\left|\Lambda\right|}{2}\right)^{3/4} \gamma ^2 
   \epsilon  \left(a^{\dagger 3}-6 N a^{\dagger}+2
   a^3\right),
\end{eqnarray}
The expectation values are, $\left< \tilde{q}_k\right>=0$, $\left< \tilde{p}_k\right>=0$. Finally, {using the inverse transformation of Eq.\eqref{Trans1}, the volume of the universe is the cube of the scale factor, $V=a^3(t)=\frac{3|\Lambda|}{8}\left(x^2+y^2\right)$ thereby,} we calculate the spectrum for the volume, at order $\gamma^2$, as the expected value of
 \begin{eqnarray}
{V(n)}&=&\left<x^2+y^2\right>\\
&=&\sqrt{\frac{3}{2\left|\Lambda\right|}} (2 n+1)+\frac{9\beta\gamma}{8}\left(\frac{3}{8\left|\Lambda\right|}\right)^{1/4}\left(\frac{4}{\sqrt{3}}-\sqrt{2\left|\Lambda\right|}\right)\left(1+2n+2n^{2}\right)\nonumber\\
     &+&\frac{3\beta^{2}\gamma^{2}}{64}\left(8-4\sqrt{6\left|\Lambda\right|}+3\left|\Lambda\right|\right)\left(6+13n+3n^{2}+2n^{3}\right).
 \end{eqnarray}
To find the GUP modified WDW--equation, we {perform a} canonical quantization {to the Hamiltonian in Eq.\eqref{HGUP}} and get
\begin{equation}
\hat{H}_{GUP}~\psi=\hat{H}_{0}\psi+i\beta\gamma\hat{H}_{1}\psi+\gamma^{2}\left(\frac{\beta^{2}}{6}+\frac{2\epsilon}{3}\right)\hat{H}_{2}\psi+2i\beta\gamma^{3}\left(\frac{\beta^{2}+2\epsilon}{3}\right)\hat{H}_{3}\psi-\gamma^{4}\left(\frac{\beta^{2}+2\epsilon}{3}\right)\hat{H}_{4}\psi=0,
\end{equation}
{where,}
\begin{eqnarray}
\hat{H}_0&=&-\frac{1}{2}\left(\frac{\partial^2}{\partial x^2}+\frac{\partial^2}{\partial y^2}\right),~~~~~~
\hat{H}_1=\left(\frac{\partial^2}{\partial x^2}+\frac{\partial^2}{\partial y^2}\right)^{3/2},~~~~~~
\hat{H}_2=\left(\frac{\partial^2}{\partial x^2}+\frac{\partial^2}{\partial y^2}\right)^{2},\nonumber\\
\hat{H}_3&=&\left(\frac{\partial^2}{\partial x^2}+\frac{\partial^2}{\partial y^2}\right)^{5/2},~~~~~~
\hat{H}_4=\left(\frac{\partial^2}{\partial x^2}+\frac{\partial^2}{\partial y^2}\right)^{3}.\label{Hs}
\end{eqnarray}
In order to simplify our analysis we will restrict to the case
$\beta=0$ and work up to the order $\gamma^{2}$.
To find the {semi--classical} effects of the {GUP} deformation we propose that the wave function is given by
$\psi=e^{i\left(S\left(x\right)+G\left(y\right)\right)}$, next we apply $\hat{H}_0$ and $\hat{H}_2$ given by Eq.\eqref{Hs}, and finally we perform a WKB--type approximation to obtain
\begin{eqnarray}
\hat{H}_{0}\psi&\approx&\frac{1}{2}\left[\left(S'\right)^{2}+\left(G'\right)^{2}\right]\psi+\frac{1}{2}\omega^{2}\left[x_{0}^{2}+y_{0}^{2}\right] \psi,\\
\hat{H}_{2}\psi&\approx& \left[ \left(G'\right)^{4}+\left(S'\right)^{4}\right]\psi+2\left[ \left(G'\right)^{2} \left(S'\right)^{2}\right]\psi,
\end{eqnarray}
where $S'=dS/dx$ and $G'=dG/dy$.
Now, 
using the usual definition
$\left(\frac{\partial S}{\partial x }\right)=P_{x}$ and $\left(\frac{\partial G}{\partial y }\right)=P_{y}$, 
we get the classical Hamiltonian 
\begin{equation}\label{H}
H=\frac{1}{2}\left({P_{x}^2+P_{y}^2}\right)-\frac{3\Lambda}{8}\left({x}^2+{y}^2\right)+\frac{2}{3}\gamma^2 \epsilon \left( P_{x}^4 +2 P_{x}^2 P_{y}^2+P_{y}^4\right).
\end{equation}
Using Hamilton's formalism we find 
the equations of motion for the coordinates $x$ and $y$
\begin{equation}
\ddot{x}-\frac{3}{4}\Lambda x-2 \gamma^2\Lambda\epsilon \left[2y \dot{x}\dot{y}+ x\left(3\dot{x}^2+\dot{y}^2\right)\right]=0,\hspace{.7cm}
\ddot{y}-\frac{3}{4}\Lambda y-2\gamma ^2\Lambda\epsilon  \left[2x\dot{y}\dot{x}+y\left(3\dot{y}^2+\dot{x}^2\right)\right]=0.
\end{equation}
Considering $\gamma^2, \epsilon\ll 1$, we can find an approximate solution to order $\gamma^{2}$,
\begin{equation}\label{yt}
x(t)=x_{0}(t)+\gamma^{2} x_{1}(t)+\mathcal{O}(\gamma^4),\hspace{.7cm}
y(t)=y_{0}(t)+\gamma^{2} y_{1}(t)+\mathcal{O}(\gamma^4).
\end{equation}
The solutions for $\Lambda>0$, to order $\gamma^{2}$ and {discarding terms of order} $\gamma^2\epsilon$, are
\begin{eqnarray}
x\left(t\right)&=&A \sinh \left(\alpha + \sqrt{\frac{3\Lambda}{4} } ~t\right)+
\gamma ^2 C \sinh \left(\xi +\sqrt{\frac{3\Lambda}{4} } ~t\right),\nonumber\\ 
y\left(t\right)&=&B \sinh \left(\beta +\sqrt{\frac{3\Lambda}{4} } ~t\right)+\gamma^2 D \sinh \left(\nu +\sqrt{\frac{3\Lambda}{4} } ~t\right).
\end{eqnarray}
To satisfy the Hamiltonian constraint $H=0$, we find $
A^2+B^2+2 \gamma ^2  \left[A C \cosh \left(\alpha -\xi \right)+B D \cosh\left(\beta -\nu\right)\right]=0
$. Taking $A=B$, $\alpha=\xi$, $\beta=\nu$ we get the volume of the Universe $V(t)=a^3(t)=x^2+y^2$,
\begin{equation}
V(t)
=B\left(B+2\gamma^2 D\right)\sinh\left(\nu-\xi\right) \sinh\left(\nu+\xi+\sqrt{3\Lambda}~ t\right).
\end{equation}
\noindent For $\Lambda<0$ 
the solutions are
\begin{eqnarray}
x\left(t\right)&=&A\sin\left(\alpha+\sqrt{\frac{3\Lambda}{4}}~t\right)+\gamma^2 C\sin\left(\xi+\sqrt{\frac{3\Lambda}{4}}~t\right),\nonumber\\
y\left(t\right)&=&B\sin\left(\beta+\sqrt{\frac{3\Lambda}{4}}~t\right)+\gamma^2 D\sin\left(\nu+\sqrt{\frac{3\Lambda}{4}}~t\right).
\end{eqnarray}
To satisfy de the deformed Hamiltonian constraint, we find that
\begin{equation}
A^2+B^2+2  \gamma^2\left[A C \cos \left(\alpha -\xi\right)+ B D \cos\left(\beta -\nu\right)\right]=0   
\end{equation}
this expression can be simplified as in the previous case, taking $A=B$, $\alpha=\xi$ y $\beta=\nu$, we get  the volume of the Universe as
\begin{equation}
V(t)=B \left(B-2 \gamma ^2 D\right) \sin (\nu -\xi ) \sin \left(\nu +\xi +\sqrt{3\Lambda }~ t\right).
\end{equation}
\begin{figure}[H]
\begin{center}
\includegraphics[width=.51\textwidth]{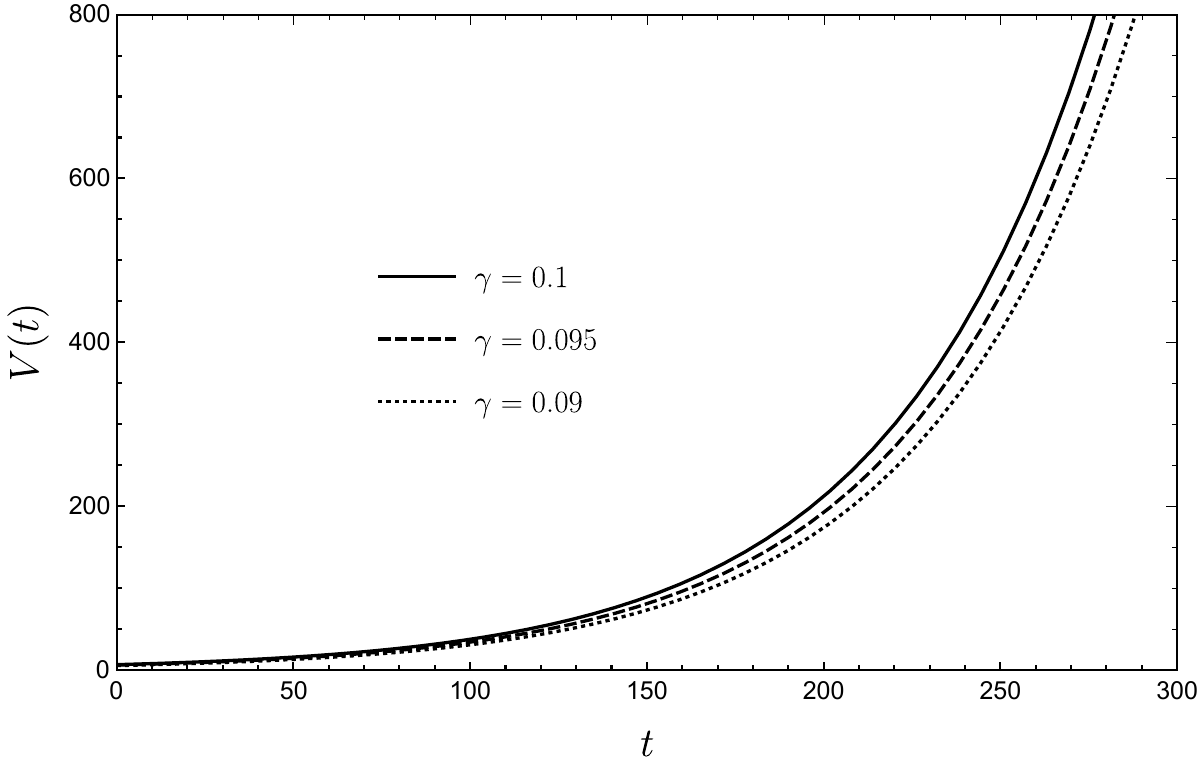}\includegraphics[width=.51\textwidth]{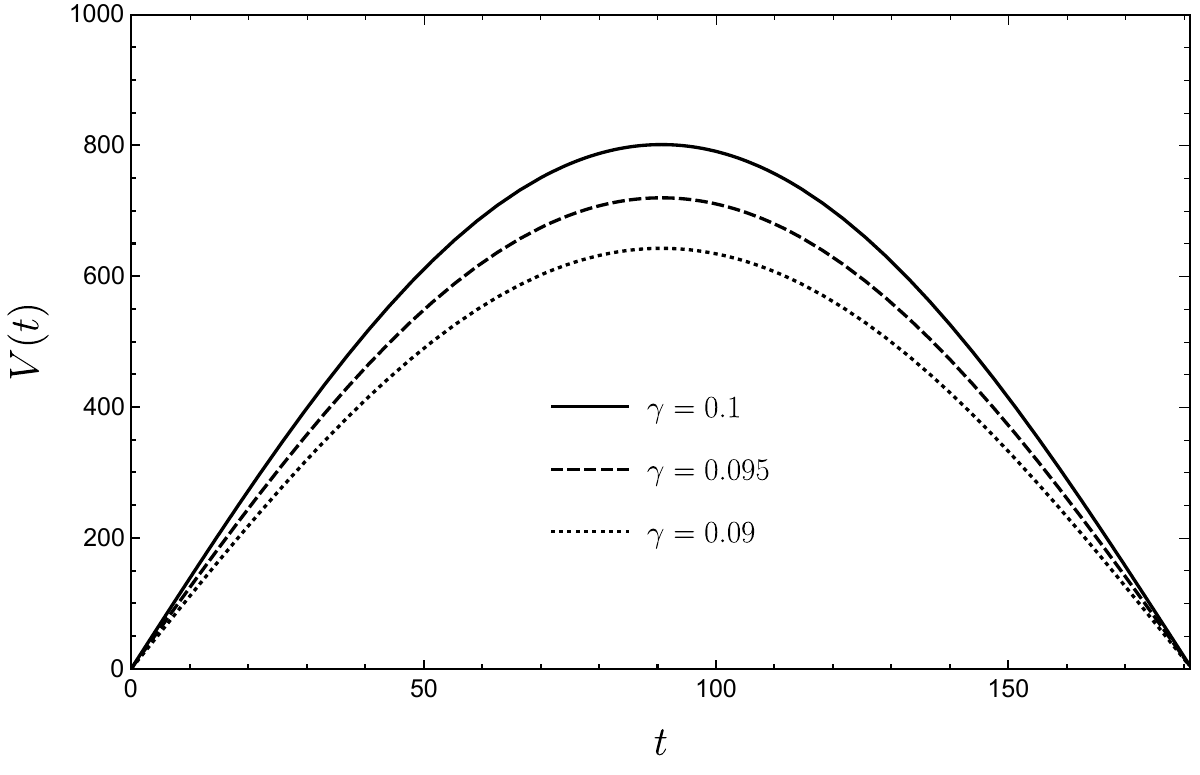}
\caption{{Plots of the classical behaviour of the scale factor. The first plot corresponds for $\Lambda>0$, we can see that the late time behavior corresponds to a de Sitter Universe, therefore the GUP effects can only be relevant at early times. The second plot is for $\Lambda<0$, we have an expanding an then contracting Universe. From the expression for the volume we see that after the contraction the volumes is negative.}}
\label{figdata}
\end{center}
\end{figure}

\section{GUP Modified Friedmann equations.}

{Introducing the GUP in the cosmological model, was simplified by transforming the Hamiltonian in Eq.(\ref{ham}) to Hamiltonian of a two dimensional oscillators. This result depends on the potential and the resulting Hamiltonian will not be the same for an arbitrary potential. For example, if we take the potential 
\begin{equation}
    V\left(\phi \right)=\frac{c_{1}}{2\mu^{2}}+\frac{ \text{b Cosh}\left(\mu\phi\right) \text{Sinh}\left(\mu\phi\right)}{\mu^{2}}+\frac{\left(c_{1}+c_{2}\right)\text{ Sinh}\left(\mu\phi\right)^{2}}{2\mu},
\end{equation}
where $b$, $c_1$, $c_2$ and $\mu$ are constants that characterize the potentials family, and use the transformation in Eq.(\ref{Trans1}), the Hamiltonian takes the simple form 
\begin{equation}
{H}=\frac{1}{2}\left(P_{y}^{2}-P_{x}^{2}\right)+\frac{c_{1}}{2}x^{2}+\frac{c_{2}}{2}y^{2}+bxy.
\end{equation}
Using the approach presented in the previous section, we apply the GUP in Eq.\eqref{q} and derive the corresponding GUP modified Hamiltonian. To first order in $\gamma^{2}$ and taking $\beta=0
$, $\epsilon=1$, the GUP modified Hamiltonian is
\begin{equation}
    {H}=\frac{1}{2}\left(P_{0y}^{2}-P_{0x}^{2}\right)+\frac{1}{4}\left(c_{1}x_{0}^{2}+c_{2}y_{0}^{2}\right)+bx_{0}y_{0}+\frac{2\gamma^{2}}{3}\left(P_{0y}^{4}-P_{0x}^{4}\right).
\end{equation}
The resulting Hamiltonian is again that of two oscillators with a perturbation. Moreover, for $b=0$ we can follow the same approach as in the previous section. We will like to point out that even if we take an arbitrary potential, (after using the transformation) the kinetic term on the Hamiltonian is the same, and the modifications will be on the potential.}

It is well known that  to study the dynamics of the Universe, we only need the Friedmann equations together with the continuity equation. Therefore, to study the GUP modifications in a model with an arbitrary potential, it is sensible construct the Friedmann equations with the GUP modifications. We start with the Lagrangian for the phantom scalar field with an arbitrary potential   
\begin{equation}
{L}=-3 a\dot{a}^{2}+a^{3}\left(\frac{\dot{\phi}^{2}}{2}-V\left(\phi\right) \right). 
\end{equation}
The Hamiltonian is given by
\begin{equation}\label{ham_phi}
    {H}=\frac{\pi_{\phi}^{2}}{2a^{3}}-\frac{\pi_{a}^{2}}{12a}+a^{3}V\left(\phi\right).
\end{equation}
From  Hamilton's equations we get the equation of motion for the scale factor $a$
\begin{equation}
    2\frac{\ddot{a}}{a}+\left(\frac{\dot{a}}{a}\right)^{2}=-\frac{\dot{\phi}^{2}}{2}+V\left(\phi\right),
\end{equation}
which is equivalent to the Friedmann equation. Moreover, the equation of motion for $\phi$, gives the Einstein--Klein--Gordon equation,
\begin{equation}
  \ddot{\phi}=\frac{d V\left(\phi\right)}{d\phi}-3\dot{\phi}H,
\end{equation}
where, as usual, $H=\dot{a}/a$. 

{To derive the GUP modified Friedmann equations, we start by applying Eq.(\ref{Trans1}) in the Hamiltonian in Eq.(\ref{ham_phi}). After applying the GUP transformation Eq.(\ref{q}) we get the GUP  Hamiltonian in the variables $(x,y)$. Finally, after applying the inverse transformation to Eq.(\ref{Trans1}), the resulting GUP Hamiltonian to order $\gamma^2$ is}
\begin{eqnarray}
{H}=a^3 V(\phi )-\frac{P_a^2}{12 a }+\frac{P_\phi ^2}{8}+\gamma ^2\frac{ e^{-5\sqrt{6} 
   \phi/2}}{108 a^7 }\left(a^{2} P_a^{2}-6 P_\phi^{2}
 \right)
    \left(2 a e^{5 \sqrt{6}
   \phi/2}+a \left(\frac{3}{8}\right)^{3/2} P_a P_\phi -\frac{3}{8} e^{\sqrt{6}\phi}\right).
\end{eqnarray}
The last step is to use  Hamilton's equation in the GUP modified Hamiltonian, to obtain the GUP modified Friedmann equation, 
\begin{align}
2\frac{\ddot{a}}{a}
&=-H^2-\frac{\dot{\phi }^2}{2}+V(\phi )+\gamma^2 \left[9a^2H^2e^{-\sqrt{6}\phi/2}\left(-H^2-\frac{\dot{\phi}^2}{2}+V(\phi)\right)-48 a^3 H^2 \left(-\frac{1}{3}H^2-\frac{\dot{\phi}^2}{2}+V(\phi)\right)\right.\\ \nonumber
&\left. +4a^8H\left(\frac{3}{8}\right)^{3/2}e^{-5\sqrt{6}\phi/2}\dot{\phi}\left(\dot{\phi }^4 +60H^2 \left(-\frac{\dot{\phi}^2}{2}+V(\phi)\right)\right)\right],
\end{align}
and the corresponding GUP modified KG equation
\begin{eqnarray}
\ddot{\phi} &=&\frac{d V\left(\phi\right)}{d\phi}-3\dot{\phi}H +\gamma^{2} \left[16\dot{\phi}^2\left(a^3H \dot{\phi}-2\frac{d V\left(\phi\right)}{d\phi}\right)+\frac{3}{2}\dot{\phi}^2 e^{-\sqrt{6}\phi/2}\left(\frac{4}{a}\frac{d V\left(\phi\right)}{d\phi}-\frac{7}{3} a^2 H\dot{\phi } \right)\right .\nonumber\\
&+&\left. a^5H\left(\frac{3}{8}\right)^{3/2}\dot{\phi}^3e^{-5\sqrt{6}\phi/2}\left(160\frac{d V\left(\phi\right)}{d\phi}-60a^3H\dot{\phi }+144\frac{a^3 H^5}{\dot{\phi }^3 } \right)\right].
\end{eqnarray}
From the modified Friedman equation, we can define an effective potential $V_{eff}$
\begin{eqnarray}
V_{eff}&=& V\left(\phi\right)+\gamma^{2} \left[9 a^2 H^2e^{-\sqrt{6}\phi/2}\left(-H^2-\frac{1}{2}\dot{\phi }^2+V(\phi)\right)-48a^3 H^2 \left(-\frac{1}{3}H^2\frac{1}{2}\dot{\phi}^2+V(\phi)\right)\right.\\ \nonumber
&+&\left.4 a^8 H \left(\frac{3}{8}\right)^{3/2} \dot{\phi }e^{-5\sqrt{6}\phi/2}\left(\dot{\phi }^4+60H^2\left(-\frac{1}{2}\dot{\phi}^2+V(\phi)\right)\right)\right].   
\end{eqnarray}
Now, we can write and effective density and pressure
\begin{equation}
    \rho^{eff}_{\phi}=\frac{\dot{\phi}^{2}}{2}+V_{eff},\quad P^{eff}_{\phi}=\frac{\dot{\phi}^{2}}{2}-V_{eff}.
\end{equation}
For $\gamma=0$, the effective potential is the potential in the action and the regular Friedmann equations are recovered.
{For general potentials, the expressions for the corrections are quite cumbersome, an as expected for second order in $\gamma$ will be more complicated. }
\section{Discussion and Outlook}
In this paper we have explored the effects of the GUP in phantom cosmology.
We exploit a transformation, that transforms the Hamiltonian of the cosmological model in to the two dimensional
harmonic oscillator. Therefore, introducing the GUP transformation can be more o less straight forward.
In particular we study the deformation  introducing modifications to the momentum.

{There is another approach where the GUP deformation is done by modifying the coordinates. This particular deformation has been in used to study the Kantowski-Sachs model \cite{pasqualeCOSMO}.} The GUP
can be satisfied by the change in coordinates $q_i=q_{0i}\left(1+\gamma^{2}P_{0}^2\right)$, $
P_{j}=P_{0j}$, where $P_0^2=P_{0j}P_{0j}$.
Substituting in Eq.\eqref{Ham1} we get the GUP  modified Hamiltonian (to order $\gamma^2$)
\begin{equation}
 \hat{H}~\psi=\left[\frac{1}{2}\left(P_{0x}^2+P_{0y}^2\right)+\frac{1}{2}\omega^2\left(x^2\left(1+2\gamma^{2}P_{0}^2+\gamma^{4}P_{0}^4\right)+y^2\left(1+2\gamma^{2}P_{0}^2+\gamma^{4}P_{0}^4\right)\right)\right]\psi=0.\nonumber
\end{equation}

{Now we obtain the classical dynamics from the quantum model\footnote{The ladder operators are given by
$
\tilde{a}=a-\frac{3}{2} \gamma ^2 \Lambda  \left(a+aN-a^{\dagger ^3}\right)$,
 $\tilde{a}^{\dagger}=a^{\dagger }-\frac{3}{2} \gamma ^2 \Lambda  \left(a^3-a^{\dagger}N+2a^{\dagger}\right)
$ and the volume  is
$a^{3}\left(n\right)=\sqrt{\frac{3}{8\Lambda}}(2 n+1).$}, using  the WKB approximation on the GUP deformed WDW--equation to obtain Hamilton-Jacobi equation.} Analysing the evolution of the Universe,  we find that the GUP deformation\footnote{The Hamiltonian constrain with $A=B$, $\alpha=\beta+\frac{\pi}{2}$ and $\nu=\beta$ is $ 4B+ \gamma ^2\left(3 B^3 \Lambda -4 C \sin (\beta-\xi )+4 D\right)=0.$} gives the volume
\begin{eqnarray}
&&V(t)=\frac{3}{8} B^2 \csc ^2(\beta -\xi )
   \sin ^2\left(\beta +\frac{1}{2}
    \sqrt{3\Lambda }
   t\right)\nonumber\\
   &&+\frac{3}{16} B \gamma ^2
   \sin \left(\beta +\frac{1}{2}
   \sqrt{3\Lambda } t\right)
   \left[
   \cot (\beta -\xi ) \csc (\beta
   -\xi ) \left(3 B^3 \Lambda 
   +4D\right) \sin \left(\xi +\frac{1}{2}
    \sqrt{3\Lambda } t\right)
    +4D\sin \left(\beta +\frac{1}{2}
    \sqrt{3\Lambda }t\right)
    \right].\nonumber
\end{eqnarray}
From the volume we can see that the GUP deformation allows for a cyclic Universe.
\begin{figure}[H]
\begin{center}
\includegraphics[width=.41\textwidth]{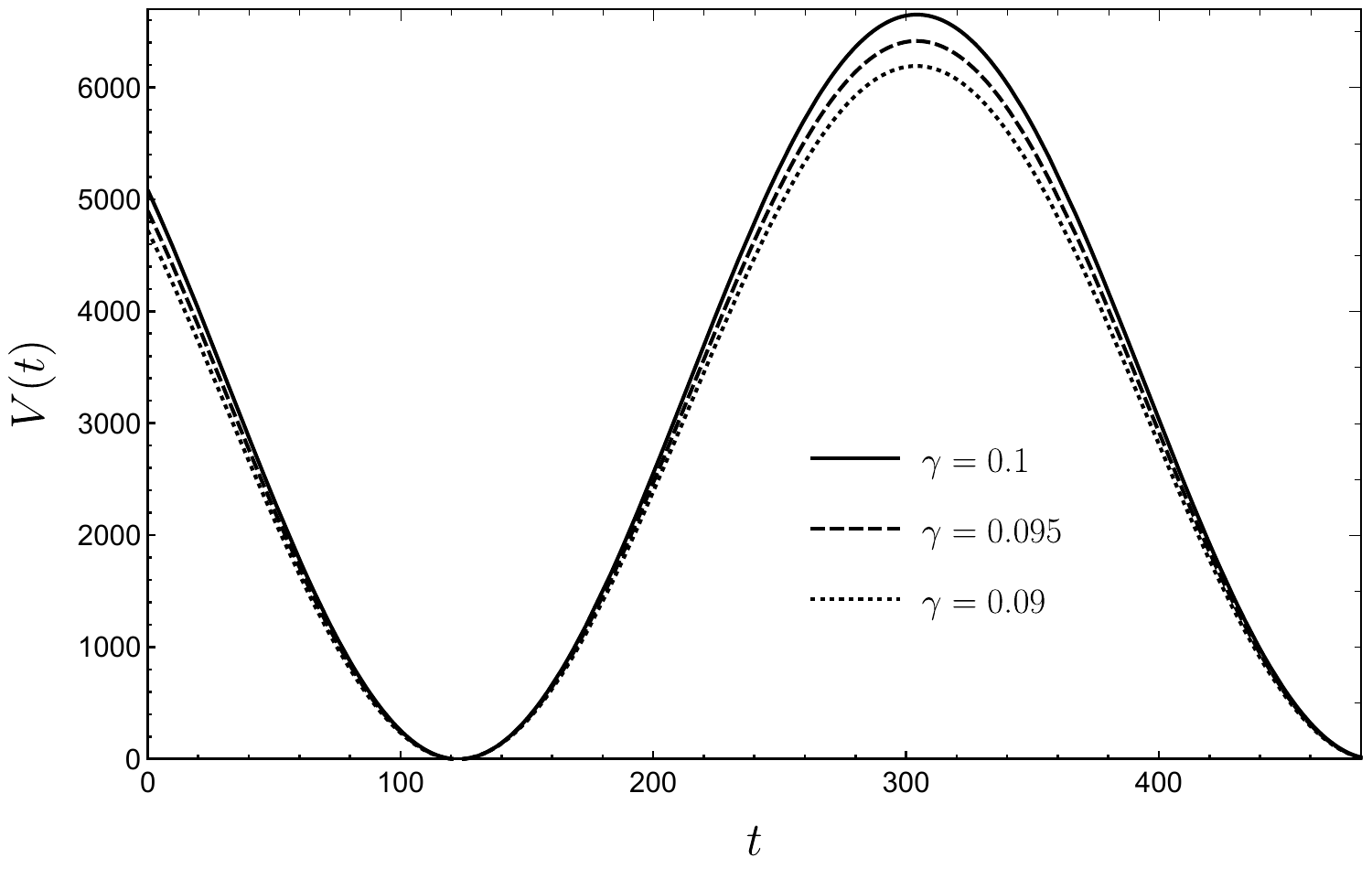}
\caption{{Plots of the classical behaviour of the scale factor. The plot is for $\Lambda<0$, we can se the effects of the perturbation allows for a cyclic Universe.}}
\label{figdata2}
\end{center}
\end{figure}
We also discussed  a more general potential \footnote{Interestingly cosh-like potentials can be phenomenologically viable, they have been related used to describe cosmological scenarios the have a united description for dark matter and dark energy \cite{tona}.} and derive the corresponding GUP modified Friedmann equations.
 Unfortunately, the resulting modifications are complicated and therefore a numerical analysis must be performed. { Moreover, we can extend this approach to introduce the GUP deformations to an arbitrary potential. By writing $V(\phi)= \Lambda + U(\phi)$, after applying the
transformation, we will get the Hamiltonian for the two dimensional oscillator plus a potential term $U(x,y)$ that is a complicated function of $x$, $y$}.
To this Hamiltonian we can introduce the GUP deformation, and following the procedure in the previous section 
we can write the modified Friedmann equations. The possibility to introduce general potentials, will allow to study the effects of GUP in the early Universe (i.e inflationary epoch) where the one can expect the GUP effects will be more relevant. These ideas are under research and will be reported elsewhere.

\section*{Acknowledgements}

{\bf J.C.L-D.} is supported by the CONACyT program  ``Apoyos complementarios para estancias sab\'aticas vinculadas a la consolidaci\'on de grupos de investigaci\'on 2022-1'' and by UAZ-2021-38339 Grant. {\bf M. S.} is supported by the grant CIIC 032/2023 and CIIC 224/2023. {\bf O. L-A.} will like to express his gratitude to CONACyT for support from the program {\it Becas nacionales para estudio de posgrado}.


\end{document}